\documentclass[twocolumn,showpacs,prd]{revtex4}

\usepackage{graphics,graphicx,amsmath}

\begin{document}

\title{Reducing eccentricity in black-hole binary evolutions with initial
  parameters from post-Newtonian inspiral} 

\author{Sascha Husa, Mark Hannam, Jos\'e A. Gonz\'alez, Ulrich Sperhake, Bernd
  Br\"ugmann} 
\affiliation{Theoretical Physics Institute, University of Jena, 07743 Jena,
  Germany} 

\begin{abstract}
Standard choices of quasi-circular orbit parameters for black-hole binary
evolutions result in eccentric inspiral. We introduce a conceptually simple method, 
which is to integrate the post-Newtonian 
equations of motion through hundreds of orbits, and read off the values of the
momenta at the separation at which we wish to start a fully general
relativistic numerical evolution. For the particular case of non-spinning
equal-mass inspiral with an initial coordinate separation of $D = 11M$ we show
that this approach reduces the eccentricity by at least a factor of five to $e
< 0.002$ as compared to using standard quasi-circular initial parameters.  
\end{abstract}


\maketitle

\section{Introduction}

Recent breakthroughs in numerical relativity
\cite{Pretorius:2005gq,Campanelli:2005dd,Baker05a} have made it possible to
accurately simulate the last orbits, merger and ringdown of a black-hole
binary system, and to compute the gravitational waves emitted in the
process. Comparison of these waveforms with those produced by analytic  
techniques (i.e., post-Newtonian [PN] methods) has already begun
\cite{Buonanno06imr,Baker:2006ha,Scheel:NRm3PN,Berti:2007fi,Pan:2007nw,Ajith:2007qp}, 
as has the process of preparing these waveforms for use in  
gravitational-wave searches~\cite{Pan:2007nw,Ajith:2007qp}.
We would like to start the comparison and template-building process
with binaries that undergo non-eccentric inspiral, but many of these simulations
model systems with non-zero eccentricity. This is ironically due to the use
of ``quasicircular orbit'' initial momenta, which are based
on an approximate circularity condition. However, a spiraling
motion must contain a radial component. We wish to find initial
tangential and radial momenta that, for a given
initial separation, lead to non-eccentric inspiral.

Pfeiffer, {\it et al.} \cite{Pfeiffer:2007} recently suggested an iterative
procedure to reduce eccentricity. They simulate a system with quasi-circular
(QC) parameters for two orbits, measure the eccentricity, make an appropriate
modification to the initial parameters (including the introduction of a radial
component to the motion), and start the simulation again. They repeat this
procedure  until the eccentricity is reduced by a factor of ten. The drawback of
their method is that it requires at least one ``false start'', which is computationally
expensive, and, as pointed out in~\cite{Pfeiffer:2007}, it is not
clear how to generalize the method to evolutions of spinning black
holes, for which the black-hole separation and wave frequency will not in
general be a monotonic function of time. We would rather find a general method
to calculate low-eccentricity parameters from the outset.

In earlier papers we used a PN approximation to calculate QC
parameters for equal-mass \cite{Bruegmann:2006at}, unequal-mass
\cite{Gonzalez2006} and spinning \cite{Gonzalez:2007hi} black holes, and 
found that they compared well with parameters calculated using more
sophisticated numerical methods \cite{Bruegmann:2006at}. In this paper
we follow up on Miller's work \cite{Miller03c} and use PN methods to estimate
initial parameters for low-eccentricity inspiral. 

Our procedure is to numerically integrate the PN equations of 
motion (at the highest PN order available; see for example
\cite{Buonanno:2005xu}) for two point particles over 
hundreds of orbits, and read off the particles' momenta when they have reached
the separation we wish to use as an initial separation in a fully general
relativistic numerical simulation. We use Mathematica for the integration,
which typically takes several seconds. Note that we could instead use
lower-order PN estimates of the radial momentum (for example by using the
quadrupole formula), but we wish to obtain the most accurate results
possible, and to use a method that can later be applied with some confidence
to spinning binaries. A PN approach was also used to introduce a
radial component to the motion in~\cite{Campanelli:2007ew}, although
the details of the method were not given.

The key to this approach is to exploit the fact that any 
initial eccentricity will decay over time due to the circularizing
effect of gravitational-wave emission, but on a time-scale of
hundreds of orbits, not the $<10$ orbits typically simulated by a
numerical code. We therefore start the PN evolution at sufficiently
large initial separation $D$ (in practice $D = 40M$) to allow 
radiation-reaction to circularize the orbits. (We will quote
time and length in units of the total initial black-hole mass $M$;
see~\cite{Bruegmann:2006at}.)

Given the parameters from PN inspiral, full GR numerical evolutions are then
performed with the BAM code 
\cite{Bruegmann:2006at,Bruegmann2004}, using the moving-puncture method
\cite{Campanelli:2005dd,Baker05a} to evolve Bowen-York initial data
\cite{Bowen80} in puncture form \cite{Brandt97b} and generated by a
pseudo-spectral method \cite{Ansorg:2004ds}. In order to perform long
evolutions with sufficient accuracy for the present purpose, it was crucial to 
modify our previous evolution algorithm to use sixth
order accurate derivative operators \cite{Husa2007a}, instead of the more
standard fourth-order accurate choice. 

We find that the PN-inspired initial momenta lead to evolutions with at least
five times less eccentricity than their QC counterparts.

In Section~\ref{sec:PNequations} we summarize the PN equations that we use and
the method to integrate them. In Section~\ref{sec:results} 
we present results from simulations of an equal-mass binary. 

\section{Integration of the PN equations of motion}
\label{sec:PNequations} 

We have used the PN equations of motion as described in
\cite{Buonanno:2005xu} in the ADMTT gauge. We have implemented both
the usual Taylor-expanded and the effective-one-body (EOB) versions of the
Hamiltonian; the calculations presented here are however all based on the
Taylor-expanded version. The PN solution in the ADMTT gauge for a 
two-body system agrees with our Bowen-York puncture initial data up to 2PN
order (see, for example, the explicit solutions in Appendix A of
\cite{Jaranowski98a}). The conservative part of the 
Hamiltonian is given up to third PN order, and was originally
derived in \cite{Jaranowski98a,Damour:2001bu,Damour:2000kk}, see also \cite{Blanchet01a,deAndrade:2000gf,Blanchet03}.
Radiation-reaction flux terms are calculated up to 3.5PN
order beyond the quadrupole order, which is achieved by averaging the
radiation flux over one orbit, assuming quasi-circular inspiral
\cite{Blanchet:1997jj,Blanchet:2001aw,Blanchet:2004ek}.  
We have also included the leading-order spin-spin and spin-orbit coupling
terms for the conservative part of the Hamiltonian
\cite{Barker1970,Barker1974,Barker1979}, and spin-induced radiation 
flux terms as described in \cite{Buonanno:2005xu} (and again averaged
over one orbit).

In the nonspinning case the PN equations of motion are a system of six coupled
ordinary  differential equations of the form \begin{eqnarray}
\frac{d x^i}{d t} & = & \frac{\partial H}{\partial p_i}, \\
\frac{d p_i}{d t} & = & - \frac{\partial H}{\partial x^i} + F_i,
\end{eqnarray} where $H$ is the PN-Hamiltonian (responsible for the
conservative part of the dynamics), $x^i$ is the separation vector between the
two particles and $p^i$ is the momentum of one particle in the center-of-mass
frame. In the spinning case the system is augmented by the evolution equations
for the spins. The quantity $F_i$ is the radiation-reaction flux term. We have
used the evolution equations precisely in the form presented in
\cite{Buonanno:2005xu}.  

\begin{figure}[h]
\includegraphics[width=8cm]{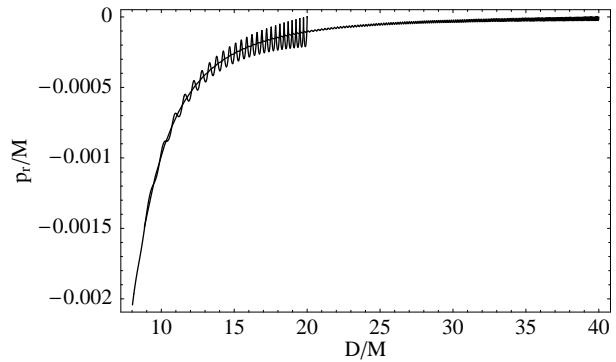}
\caption{\label{fig:eccentricity_decay} The radial momentum component is
  plotted versus separation 
 for PN-inspirals starting from $D=20M$ and $D=40M$.
  A separation of $D=20M$ is clearly not sufficient to produce non-eccentric
  inspiral parameters,  
  since  small oscillations can still be seen at $D=11M$, while for $D=40M$
  the initial eccentricity has essentially decayed away.}
\end{figure}

Starting at a suitably large initial separation ($D = 40M$ is used in
practice for an equal-mass binary), initial momenta are chosen using the 3PN
formula given in \cite{Bruegmann:2006at}. 
We have checked that $D = 20M$ would be somewhat too close --- 
small oscillations in the radius are still visible at $D=11M$, where we wish
to start the evolution of the full Einstein equations; see
Fig.~(\ref{fig:eccentricity_decay}). Integrating the PN equations from
$D = 100 M$ (2071.5 orbits) makes very little difference for the inspiral
parameters. For the full numerical evolution to start at $D=11M$ the
tangential component of the momentum would change by $4\times 10^{-4}\%$, the
radial component by $0.2\%$ as compared to using a PN-inspiral from $D =
40M$.

The equations are then integrated in Mathematica 5.2 using the {\tt NDSolve}
function with different options for the integration algorithm, tolerance
levels and internal precision to check the accuracy of the
results. Mathematica stops the integration automatically when the PN equations
of motion become ill-defined. Several consistency checks are applied to make
sure the correct equations are integrated: When radiation reaction is switched
off, energy and angular momentum should be constant up to numerical error. To
check this, we integrate the conservative equations of motion with initial 
separations $40M$ and $50M$ for about 200 orbits
and monitor the relative decay of energy and angular momentum,
which remain below $3.5\times 10^{-6}$ for the {\tt NDSolve} options we have
used for the results presented here. When radiation reaction is switched on,
the PN equations of motion imply the identity  
$$
\frac{dE}{dt} = \frac{d\varphi}{dt} 
     {\boldsymbol \lambda} \cdot \left\langle \frac{d {\boldsymbol J}}{dt} \right\rangle,
$$
in the circular equal-mass case \cite{Buonanno:2005xu}, where 
$\varphi$ is the orbital phase, ${\boldsymbol \lambda}$ the time independent 
unit vector in the direction of the orbital
angular momentum, and $\langle \cdot \rangle$ denotes the orbital average.
We have checked that for the actual inspiral (which is not exactly circular)
this equation (without taking the orbital average) is satisfied to better than $2\times 10^{-4}$ over the entire
inspiral. 

The system displays some initial eccentricity, but this decays and by the time
the particles are at a separation suitable for a numerical evolution (i.e., $D <
20M$), the inspiral has negligible eccentricity, as shown in
Fig.~(\ref{fig:eccentricity_decay}). Similar plots are also shown in
\cite{Miller03c}. 

We now wish to perform a full GR numerical evolution of the last orbits of the binary
system. The puncture initial data solver requires as input the black hole's
masses, positions, and momenta. Given the masses and some desired initial
separation, we can read off the appropriate momenta from the integrated
solution $(x^i(t), p^i(t))$ of the PN equations of motion.

\section{Numerical results} 
\label{sec:results}

We consider the configurations shown in Table~\ref{tab:configurations}. The
black-hole punctures are placed at an initial separation of $D = 11M$. They
are given either quasi-circular orbit (QC) parameters as estimated from the
2PN-accurate expression in \cite{Kidder:1995zr,Bruegmann:2007a}, or the
PN-inspired low-eccentricity parameters described in
Section~\ref{sec:PNequations}. When evolved both initial configurations lead
to about seven orbits before merger.  

\begin{table}
\begin{ruledtabular}
\begin{tabular}{l|r|r|r|r|r|r}
Configuration  & $P_x/M$  & $P_y/M$                     & $e_D$  & $e_{\omega}$ \\
 \hline
QC11 	& $\mp0.0899395$  &  0                          & 0.012   & 0.01    \\
E11	& $\mp0.0900993$  & $\mp7.09412 \times 10^{-4}$ & 0.002   & 0.002   \\
\end{tabular}
\end{ruledtabular}
\caption{Initial physical parameters for a standard ``quasi-circular orbit''
  (QC11) and PN-inspired low-eccentricity (E11) configuration. Both have an
  initial coordinate separation of $D = 11M$ and the punctures are placed at
  $y = \pm 5.5M$. The initial eccentricities $e_D$ and $e_{\omega}$ are
  estimated using Eqns.~(\ref{eqn:eD}) and (\ref{eqn:eOmega}).
\label{tab:configurations}}
\end{table}

In the notation of \cite{Bruegmann:2006at}, the data were evolved with a grid
setup  of $\chi_{\eta=2}[5\times 64:5\times  128:6]$ using the sixth-order
accurate spatial finite differencing stencils, as described in detail in
\cite{Husa2007a}. Lower-resolution runs and convergence tests show that the
simulations are cleanly sixth-order convergent up to around $1000M$ of
evolution time and drop slightly in convergence order after that. In this
paper we only need the simulation up to $t = 1000M$, at which
time the uncertainty in $D(t)$  is 0.6\%; for all earlier times it is lower. 

\begin{center}
\begin{figure}[h]
\vspace{0.5cm}
\includegraphics[width=8cm]{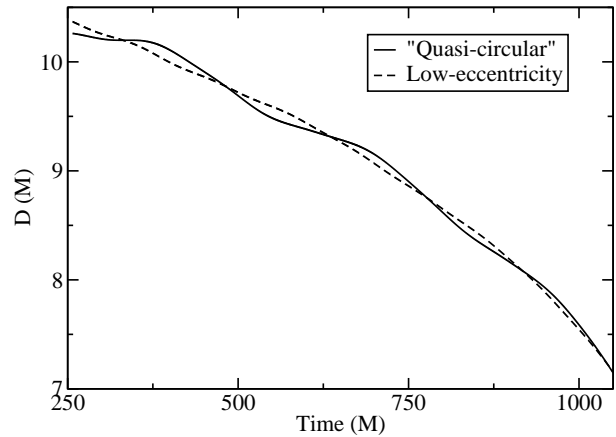}
\caption{\label{fig:D11_separation} Coordinate separation of the punctures as a
  function of time for the quasi-circular (QC11) and PN-inspired
  low-eccentricity (E11) initial parameters.} 
\end{figure}
\end{center}

Figure~\ref{fig:D11_separation} shows the coordinate separation of the
punctures as a function of time, for simulations with the QC11 and E11 initial 
parameters. The figure begins at $t = 257M$,
the time at which the binary completes one orbit. Before that time there are
oscillations in $D(t)$ due to gauge adjustments; the initially stationary
punctures pick up speed, the lapse and shift adapt to the dynamical gauge
conditions, and the numerical grid points rapidly retract from the extra
asymptotically flat ends in the puncture initial data \cite{Hannam:2006vv,
Hannam:2006xw}. All of these effects preclude a meaningful estimate of the
eccentricity during the first orbit. The figure ends at $t = 1050M$, when the
system has completed a further four orbits. We can clearly see oscillations
due to eccentricity in the QC11 data, while the E11 data appears relatively
eccentricity-free. 

We use two methods to estimate the eccentricity, as also used in
\cite{Baker:2006ha,Buonanno06imr,Pfeiffer:2007}. Assume that we know the
zero-eccentricity quasi-circular inspiral for our system, and denote the
corresponding coordinate separation of the punctures as a function of time as
$D_c(t)$ and the orbital frequency as $\omega_c(t)$. The coordinate separation
and orbital frequency for any given numerical evolution are $D(t)$ and
$\omega(t)$. The eccentricity can be estimated by extrema in either
\begin{equation}  
e_D(t) = \frac{D(t) - D_c(t) }{D_c(t)}, \label{eqn:eD}
\end{equation} or \begin{equation}
e_{\omega}(t) = \frac{\omega(t) - \omega_c(t)}{2 \omega_c(t)}.  \label{eqn:eOmega}
\end{equation}

In practice we estimate $D_c(t)$ by fitting a curve through the numerical
$D(t)$ for the low-eccentricity simulation \begin{equation}
D_c(t) = a T^{1/2} + b T + c T^{3/2} + d T^2,
\end{equation} where $T = T_M - t$ and $T_M$ is a rough estimate of the merger
time. For the E11 simulation, we choose $t_M = 1270M$. Similarly we follow
\cite{Baker:2006ha} and fit a fourth-order polynomial 
in time through the $\omega(t)$ curve for the low-eccentricity simulation and
obtain $\omega_c(t)$.

For the quasi-circular simulation QC11, Eqn.~(\ref{eqn:eD}) gives $e_D = 0.012
\pm 0.002$ and Eqn.~(\ref{eqn:eOmega}) gives $e_{\omega} = 0.01 \pm
0.001$. The uncertainties are estimated by repeating the calculation with
curves $D_c(t)$ and $\omega_c(t)$ fit through the eccentric QC11 data. Note
that the frequency method (\ref{eqn:eOmega}) gives a lower value than the
separation method (\ref{eqn:eD}); similar results were found in
\cite{Buonanno06imr,Pfeiffer:2007}. 

For the low-eccentricity E11 simulation, we find $e_D = 0.002 \pm 0.001$ and
$e_{\omega} = 0.002 \pm 0.0005$. The large uncertainty in the value from the
separation method is due to the larger uncertainty in the curve fit through
$D(t)$. For both estimates, however, the one firm conclusion we can draw is
that the eccentricity in the E11 simulation is significantly lower (by a
factor of at least five or six) than that for the QC11 simulation.

\begin{center}
\begin{figure}[h]
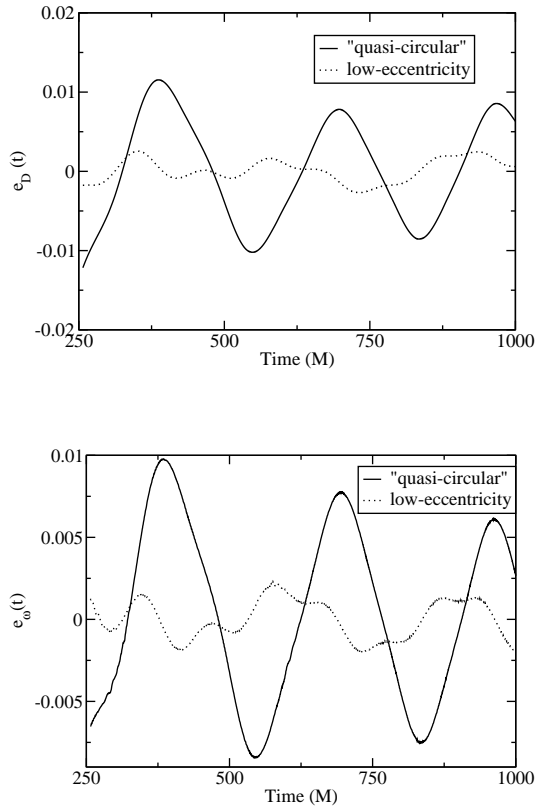

\includegraphics[width=7cm]{figures/eD.eps}\\[1cm]
\includegraphics[width=7cm]{figures/eOm.eps}
\caption{\label{fig:eccentricity} The functions $e_D(t)$ and $e_{\omega}(t)$
  for the QC11 and E11 simulations. The extrema of these functions give an
  estimate of the eccentricity, as described in the text.} 
\end{figure}
\end{center}

The functions $e_D(t)$ and $e_{\omega}(t)$ are shown in
Figure~\ref{fig:eccentricity}. For the QC11 simulations we see clear
oscillations due to the eccentricity. The curves for the E11 run are much
noisier. This may be due to errors in the curve fit through the E11 $D(t)$ and 
$\omega(t)$ being of a similar magnitude to the oscillations due to the
remaining eccentricity.

For convenience, we have computed analytical fits for the  
momentum parameters from our numerical solution to the PN evolution
equations. These fits have been computed from a PN-inspiral starting at $D=100
M$. With the expressions
\begin{eqnarray*}
\dot r &=& -2.7069 \left( 1.21329 - \frac{1.5053}{\sqrt r} + \frac{2.60155}{r} \right) r^{-2.993} \\
p_r    &=& -1.9188 \left( 1.76084 - \frac{5.3029}{\sqrt r} + \frac{9.06417}{r} \right) r^{-3.288}\\
p_t    &=& \pm \left( P_{3PN}(r) - \frac{35.0988}{r^{5.36702}} \right)
\end{eqnarray*}
the relative errors with respect to our numerical results are smaller than
$0.3 \%$ for $\dot r$ and $p_r$ and $3.5 \times 10^{-4}$ for $p_t$ over the
range $D = 8M$ to $D = 20M$.  Here $P_{3PN}(r)$ is the 3PN-accurate quasi-circular
value, which we have used previously in \cite{Bruegmann:2006at}. In Table
(\ref{tab:numbers}) we tabulate results for selected values of the black-hole
separation.  
\begin{table}
\begin{ruledtabular}
\begin{tabular}{l|r|r|r}
separation/M & $-\dot r$ ($\times 10^{-3}$)  & $-p_r/M$ ($\times 10^{-3}$)  & $p_t/M$ \\
 \hline
8.0    &   5.3857   &   2.0906   &   0.112349   \\
8.5    &   4.4944   &   1.7023   &   0.107614   \\
9.0    &   3.7839   &   1.4019   &   0.103376   \\
9.5    &   3.2133   &   1.1670   &   0.099561   \\
10.0   &   2.7512   &   0.9813   &   0.096109   \\
10.5   &   2.3736   &   0.8328   &   0.092968   \\
11.0   &   2.0624   &   0.7128   &   0.090099   \\
11.5   &   1.8039   &   0.6150   &   0.087464   \\
12.0   &   1.5872   &   0.5343   &   0.085037   \\
12.5   &   1.4042   &   0.4672   &   0.082791   \\
13.0   &   1.2487   &   0.4110   &   0.080706   \\
13.5   &   1.1156   &   0.3635   &   0.078765   \\
14.0   &   1.0010   &   0.3231   &   0.076952   \\
14.5   &   0.9018   &   0.2886   &   0.075255   \\
15.0   &   0.8155   &   0.2589   &   0.073661   \\
15.5   &   0.7398   &   0.2331   &   0.072161   \\
16.0   &   0.6734   &   0.2106   &   0.070746   \\
16.5   &   0.6150   &   0.1911   &   0.069409   \\
17.0   &   0.5629   &   0.1738   &   0.068143   \\
17.5   &   0.5169   &   0.1586   &   0.066942   \\
18.0   &   0.4757   &   0.1452   &   0.065801   \\
18.5   &   0.4386   &   0.1331   &   0.064715   \\
19.0   &   0.4055   &   0.1224   &   0.063679   \\
19.5   &   0.3757   &   0.1129   &   0.062691   \\
20.0   &   0.3488   &   0.1043   &   0.061747   \\
\end{tabular}
\end{ruledtabular}
\caption{Radial velocity and radial ($p_r$) and tangential ($p_t$)
components of the black hole momentum as a function
of the separation in ADMTT coordinates for selected values of the separation.
The numbers have been produced from a PN-inspiral from $D=100 M$.
\label{tab:numbers}}
\end{table}

\section{Conclusions}

We have presented a conceptually simple method to specify very low eccentricity
initial-data parameters for the numerical evolution of binary
systems. We integrate PN equations of motion from an initial
separation of $D = 40M$ to the separation we wish to use as
the starting point for a numerical evolution of the full Einstein
equations. Initial conditions for the PN inspiral are taken from a 3PN
accurate circular orbit condition. These conditions lead to a small
initial eccentricity that radiates away by the time we read off the
parameters for our full GR numerical evolution. 

The PN equations are accurate to 3PN order in the conservative
part when neglecting spins. For spin-spin and spin-orbit interactions we have
only implemented the leading order terms, although these were not used in the
application presented here. Radiation-reaction is implemented via averaging
over orbits, and is accurate to 3.5PN order beyond the leading quadrupole
contribution. The method can in principle be applied to general black-hole
initial data, and applications to unequal-mass and spinning cases will be
presented elsewhere.  

We report in detail on an equal-mass inspiral, where the evolution of the full
Einstein equations is started at a coordinate separation of $D = 11M$
\cite{spinaltap}. Our
method reduces the eccentricity by at least a factor of five to a value below
$e = 0.002$. Remaining oscillations in the coordinate distance of the two
black holes cannot clearly be identified as eccentricity. We also provide a
curve fit and table of low-eccentricity parameters for equal-mass
binaries. 

An important corollary from the success of this method is that the input
parameters in the initial data construction ($\{m_i,p_i,D\}$ in the
Bowen-York extrinsic curvature and conformal flatness ansatz) actually
correspond to the physical properties of the black holes during a long-term
evolution with excellent accuracy. 
One might instead have found that the presence of ``junk''
radiation spoils the initial data, and that after this radiation has
left the system the dynamics of the black holes are very different
from what one would have expected from, for example, a PN evolution
with the same physical parameters. The fact that PN low-eccentricity
parameters translate to low-eccentricity numerical evolutions with
Bowen-York puncture data suggest that neither the junk radiation nor the 
constraint-solution procedure adversely affect the physics of the system.

Pfeiffer, {\it et al.}~\cite{Pfeiffer:2007} have shown that waveforms from
quasi-circular and low-eccentricity parameters have large fitting factors ($>
0.99$ for $l\leq 4$ multipole contributions), and conclude that
``quasi-circular'' waveforms will be sufficiently useful for gravitational-wave
detection.  However, the use of waveforms with the lowest
eccentricity possible will be necessary to make the most accurate matching
possible to PN inspiral waveforms. Low-eccentricity waveforms
were compared  with PN waveforms in \cite{Baker:2006ha}, and we
will give a further detailed comparison in \cite{Hannam:2007}.

\acknowledgments
This work was supported in part by
DFG grant SFB/Transregio~7 ``Gravitational Wave Astronomy''.
We thank the DEISA Consortium (co-funded by the EU, FP6 project
508830), for support within the DEISA Extreme Computing Initiative
(www.deisa.org).
Computations were performed at LRZ (Munich) and at our
in-house Linux clusters Doppler and Kepler.

\bibliography{refs}

\end{document}